\documentclass[10pt]{iopart}
\usepackage[utf8]{inputenc}
\usepackage[english]{babel}
\usepackage{amsmath}
\usepackage{graphicx}
\usepackage{xcolor}
\usepackage{ulem}     
\usepackage{caption}
\usepackage{subcaption}
\RequirePackage[hyphens]{url}

\renewcommand{\sout}[1]{}
\providecommand{\db}[1]{}
\providecommand{\ue}[1]{#1}
\providecommand{\jt}[1]{#1}

\begin{document}

\title{
Macroscopic parameterization of positive streamer heads in air}

\author{Dennis Bouwman$^1$, Jannis Teunissen$^1$, and Ute Ebert$^{1,2}$}
\address{$^1$ Multiscale Dynamics Group, Centrum Wiskunde \& Informatica (CWI), Amsterdam, The Netherlands}
\address{$^2$ Department of Applied Physics, Eindhoven University of Technology, Eindhoven, The Netherlands}

\begin{abstract}
    The growth of streamer discharges is determined at their heads, for individual streamers as well as in collective phenomena, such as streamer trees or coronas or streamer bursts ahead of lightning leaders.
    Some properties of the streamer heads, such as velocity $v$ and radius $R$ now can be measured quite well, but this is very challenging for others such as the maximal electric field, the charge content of the streamer head and the degree of chemical excitation and ionization in the streamer channel.
    Here we develop, test and evaluate a macroscopic approximation for positive streamer heads in air that relates macroscopic streamer head properties to each other.
    In particular, we find that velocity $v$, radius $R$ and background field $E_{\rm bg}$ determine the complete profile of streamer heads with photoionization, if they propagate steadily.
    We also \sout{revisit} \ue{review} Naidis' approximate relation between $v$, $R$ and the maximal field $E_{\rm max}$.
    The approximate head model \ue{developed in the present paper} consists of three first-order ordinary differential equations along the streamer axis.
    It is derived from the classical fluid model for streamer discharges by assuming axisymmetry, steady streamer propagation (\ue{i.e.}\ with constant velocity and shape), and a \sout{(semi-)}spherical shape of the charge layer \ue{around the streamer head}.
    The \ue{new reduced} model \sout{shows good agreement} \ue{agrees well} with full solutions of the classical fluid model, even when it is applied to accelerating streamers.
    Therefore the model can be used for evaluations of experiments, like \ue{for the determination} \sout{measurements} of the maximal electric field \ue{from radius and velocity of the streamer}. It is also a step towards constructing reduced models for the collective dynamics of \sout{many} \ue{multi-}streamer discharges.
\end{abstract}


Published in {\it Plasma Sources Sci.\ Technol.} {\bf 34} (2025) 025015 on Feb.\ 25, 2025, \\
submitted Sept.\ 6, 2024, revised Nov.\ 20, 2024, 
accepted Jan.\ 28, 2025.

\maketitle



\ioptwocol

\section{Introduction}  \label{sec:intro}

\subsection{A streamer head model as a building block for quantitative multi-streamer models}

Streamer discharges commonly appear when non-ionized gas is exposed to sufficiently high electric fields~\cite{Nijdam_2020}. When a large gas volume is exposed to a sufficiently high voltage, streamers appear in large numbers in trees or bursts or coronas. Illustrative sequences of images of the dynamics can be found, e.g., for sprite discharges in the mesosphere \cite{cummerSubmillisecondImagingSprite2006, kanmaeDiameterspeedRelationSprite2012, stenbaek-nielsenHighSpeedObservationsSprite2013a} or for lab discharges in ambient air where pulsed voltages of megavolt amplitude are applied \cite{kochkinExperimentalStudyHard2012,huiskampSubNanosecondTransient2017}. Here hundreds of streamers --- or probably many more, as fast imaging is limited by the camera resolution --- can be seen to propagate simultaneously. The path of each streamer is paved by a streamer head that possesses an intricate inner structure, as described \sout{below} \ue{in Section~\ref{sec:1.3}}.  

Fully three-dimensional simulations of micro-physics based fluid and particle models have made much progress recently \cite{Nijdam2016, teunissen_simulating_2017, wang2023, marskarStochasticSelfconsistent3D2024}; they resolve the electron and ion densities and the electric field. 
The dynamics of one or a handful streamers can now be simulated with such models, and the results for positive streamers in air have been bench-marked against numerical implementations of other groups~\cite{bagheriComparisonSixSimulation2018} and successfully compared with experiments~\cite{Nijdam2016, li2021, wang2023}. But more complex dynamics of many streamers or over much longer times are outside the reach of current fluid models. 

\ue{To construct a systematic approximation, we here} \sout{Here we} analyze steadily propagating streamer heads (i.e., with constant velocity and shape, see~\cite{Qin_2014a, Francisco2021-no1, Li_2022}) simulated with the classical fluid streamer model, and we develop a macroscopic approximation of the underlying physics. In the resulting reduced model the streamer head is characterized by macroscopic parameters like velocity $v$, radius $R$, maximal field $E_{\rm max}$ and excitation or ionization density $n_e$ within a given background electric field ${E_{\rm bg}}$; this head parametrization can at a later stage be \ue{extended to non-steady propagation and} implemented into a multi-streamer model for growing trees of linear conductors. The classical version of this type of reduced models is the dielectric breakdown model (DBM)~\cite{niemeyerFractalDimensionDielectric1984} with numerous versions developed in different fields of science. We refer, in particular, to the model with finite conductivity of the streamer channels as suggested in~\cite{luqueGrowingDischargeTrees2014}, but taking now the enormous variability of streamer head dynamics into account~\cite{brielsPositiveNegativeStreamers2008,Nijdam_2020}.

In the present paper we study positive streamer discharges in artificial air (80\% N$_2$, 20\% O$_2$) at 1 bar and 300 K. We study the properties of streamer heads \sout{both with} \ue{including either} photo-ionization \sout{and with} \ue{or}  background ionization as a source of free electrons ahead of the streamers.

\subsection{Evaluation of experiments with the help of a streamer head model}

The electric field profile of a streamer determines the concentrations of chemical products in the trace of the discharges~\cite{bouwmanEstimating2023}, but \sout{it} \ue{this field} is very challenging to measure~\cite{dogariuSpeciesIndependentFemtosecondLocalized2017,chngElectricFieldMeasurements2020}.
A second motivation to analyze streamer heads is therefore to obtain information about their electric field profile from more easily measurable parameters such as radius and velocity.
A much-quoted paper by Naidis~\cite{naidisPositiveNegativeStreamers2009} suggests a mathematical relation between velocity $v$, radius $R$ and maximal electric field $E_{\rm max}$ at a streamer head
$E_{\rm max} = E_{\rm max}(v,R)$.
This means that the two relatively easily measurable quantities $v$ and $R$ determine $E_{\rm max}$.
In this paper, we will derive a more systematic approximation for the relation between $v$, $R$ and $E_{\rm max}$, \ue{and we will show that this relation also depends} \sout{and its dependence} on the background field $E_{\rm bg}$, and on background ionization or photoionization:
\begin{equation}\label{eq:Emax}
    E_{\rm max} = E_{\rm max}(v,R,E_{\rm bg}).
\end{equation}
We also discuss other modeling-based relations between macroscopic streamer parameters, thereby extending on our previous work~\cite{bouwmanEstimating2023}.

\subsection{Streamer regions and macroscopic parameters} \label{sec:1.3}

Our model reduction is based on identifying different regions in the discharge.
As shown in numerous publications, e.g., recently in Figures 1 - 4 of our paper~\cite{bouwmanEstimating2023} for a steadily propagating streamer, a streamer consists of several regions that can be characterized by different macroscopic parameters:
\begin{itemize}
    \item[1.] {\bf The background} is characterized by two parameters: by the background field $E_{\rm bg}$ that here is assumed to be constant, and either by the background electron density $n_0$ (which is easier to treat) or by photo-ionization emitted from the streamer. The relevant ionization density is here the electron density, as motion and reactions of electrons drive plasma growth and front dynamics.
    \item[2.] {\bf In the avalanche zone} the electric field is above the breakdown value $E_k$ due to the approaching curved streamer head. Here electrons multiply and drift toward the streamer head, but the density of electrons and ions is still small enough for charge effects to be negligible.
    \item[3.] {\bf The charge layer} forms where \ue{electron} and ion density are high enough for space charge effects and to screen the electric field behind it. The layer can be parameterized by velocity $v$, radius $R$, maximal field $E_{\rm max}$, layer width $\ell$ and a measure of the charge content $Q$ of the head. The charge layer needs to be thin ($\ell \ll R$) for the characteristic field enhancement at the streamer tip. 
    All these quantities will be defined in detail further below. 
    \item[4.] {\bf The interior} of the streamer head is parameterized by an ionization density $n_{\rm ch}$ which we define as the maximum of the ion density $n_+$. This maximum is reached at $z_{\rm ch}$ where the electric field has decreased to the breakdown value $E(z_{\rm ch})=E_k$ behind the charge layer. 
    Here the plasma is not yet neutral, $n_e<n_{\rm ch}$, and the electric field still varies. 
    Note that the most relevant parameter here is the deposited ion charge density $n_{\rm ch}$, and the corresponding density of molecular excitations and dissociations. $n_{\rm ch}$ also determines the later electron density when the streamer plasma channel is approaching electric neutrality. On the contrary, the electron density $n_e$ is the most relevant parameter in the avalanche zone.
\end{itemize}

\subsection{Approach and content of the paper}

The starting point of the investigation is the classical fluid model for positive streamers in air at 1~bar and 300~K. The model is expressed by a set of partial differential equations (PDEs), and it here will be called the PDE model. 

In section~\ref{sec:ODE}, we first introduce the PDE model. Then we assume that the streamer head propagates steadily, i.e., with constant velocity and shape, and we transform the PDE model to a frame moving with the streamer. Furthermore previous simulation results of the PDE model show that the charge layer \ue{at the front part of the streamer head} is approximated quite well by a \sout{(semi-)}spherical shape~\cite{bouwmanEstimating2023}. Using an ansatz that implements this shape, the streamer dynamics can be expressed by a set of three ordinary differential equations (ODEs) of first order on the streamer axis. Our ODE-model therefore could be called a 1.5D model, i.e., an effectively 1D model that implements off-axis parameters like the streamer radius implicitly. 

We then show that a steady streamer head solution in the ODE-model is uniquely determined by four parameters. They can be chosen as the two internal parameters of radius $R$ and velocity $v$, plus the two external parameters of background field $E_{\rm bg}$ and background ionization $n_0$ in the case of a streamer with background ionization, but without photoionization. Later, in section~\ref{sec:photo}, photoionization is used. In this case only three parameters, like  $(v,\,R,\,E_{\rm bg})$, are needed to define a unique steady head solution. 

In passing, we remark that in contrast to some more heuristic approximations in our earlier paper~\cite{bouwmanEstimating2023}, the curvature of the space charge layer is now fully taken into account, and the electric field profile is calculated self-consistently. 

In section~\ref{sec:Sec3} solutions of the ODE approximation with background ionization are discussed, and how they depend on external and internal parameters. 
Section~\ref{sec:photo} treats streamers with photoionization. The photon source is here approximated as a point source on the axis of the streamer head, and the steady head solutions are calculated  with an iterative scheme. 

Section~\ref{sec:res_Sph} shows that the ODE approximation with photoionization agrees well with steady solutions of the PDE-model for a range of parameters. Therefore the ODE-approximation can be used for quantitative analysis. In particular, the dependence of the maximal electric field on radius and velocity is studied in section~\ref{sec:Emax}, with a central result presented in figure~\ref{fig:vR-relation-Ebg}. The figure can be used to extract the maximal field $E_{\rm max}$ from experiments when $v$ and $R$ are measured.

In section~\ref{sec:acc} we take a first step beyond the assumption of steady propagation. PDE solutions of accelerating streamers with photo-ionization are presented and the instantaneous $E_{\rm max}\ue{(t)}$ is predicted from $v\ue{(t)}$, $R\ue{(t)}$ and $E_{\rm bg}$ \ue{using the steady ODE-approximation} \sout{ODE-relation} (\ref{eq:Emax}). The agreement is encouraging.

We end with conclusions, and with an outlook on future steps in multi-streamer modeling and in experimental evaluations, and with a discussion of the multiplicity of ODE-solutions, and of the dynamic selection of steady head solutions.

\newpage

\section{Construction of the ODE-approximation}  \label{sec:ODE}

\subsection{Definition of the classical fluid model}
\label{sec:fluid}

The starting point of the present paper is the classical fluid model for streamer discharges in air, to be called the PDE model below:
\begin{eqnarray} \label{eq:model1}
    \partial_t n_e &=& \nabla\cdot\left( \mu{\bf E} \,n_e + D \nabla n_e\right) + \bar{\alpha} \,\mu E \,n_e + S,\\
    \partial_t n_+ &=& \bar{\alpha}\,\mu E\, n_e +S, \label{eq:model2a}\\
    \nabla\cdot {\bf E} &=&\frac\rho {\epsilon_0}, \quad \rho = e\,(n_+-n_e) \label{eq:model3}.
\end{eqnarray}
Here $n_e$ is the electron number density, $n_+$ the number density of positive minus negative ions, $\rho$ the charge density, $D$ is the electron diffusion constant, 
$\mu$ is the electron mobility, $-\mu {\bf E}$ is the local electron drift velocity, $\epsilon_0$ is the vacuum permittivity, $e$ is the elementary charge, $\bar{\alpha} = \alpha - \eta$ is the effective ionization coefficient, i.e., the ionization coefficient $\alpha$ minus the attachment coefficient $\eta$, and $S$ represents additional source terms.
The transport and reaction coefficients $D$, $\mu$ and $\bar{\alpha}$ are functions of the local electric field strength $E = |{\bf E}|$.
They were computed from Phelps cross sections~\cite{phelps_anisotropic_1985} for N$_2$ and O$_2$ using Bolsig+ with the temporal growth model, as discussed in~\cite{Li_2022}. The numerical values of $\mu(E)$ and $\bar{\alpha}(E)$ are  available in the supplement https://doi.org/10.5281/zenodo.14192466. 
The critical field $E_k$ is defined as the field strength at which $\bar{\alpha} = 0$.

In this paper the source term $S$ will either be zero or it will represent photoionization, as discussed in section~\ref{sec:photo}.
We remark that a PDE model can contain additional species and reactions that also contribute to $S$, such as electron detachment from a negative ion.
However, for simplicity of the presentation, we here discuss the simplest \ue{and classical} case of only two charged species: a density $n_e$ of mobile electrons and a density $n_+$ of immobile positive minus negative ions.


\subsection{Equations on the axis in a co-moving frame}

We will now transform the classical fluid model to a coordinate system moving along with the streamer head, making the following assumptions:
\begin{itemize}
  \item The streamer is axisymmetric, propagating with a velocity $v$ into the positive $z$ direction.
  \item The effects of electron diffusion can be neglected.
  \item The propagation is steady, i.e., the velocity $v$ is constant and other streamer properties do not change in time in a co-moving frame.
\end{itemize}
On the symmetry axis, the expression $\nabla\cdot\left( \mu {\bf E}\, n_e \right)$ from equation~\eqref{eq:model1} can be written as
\begin{eqnarray} \label{eq:nabla}
\nabla\cdot\left( \mu {\bf E}\, n_e \right) &= \mu \mathbf{E} \cdot \nabla n_e + n_e \mathbf{E} \cdot \nabla \mu + \mu n_e \nabla \cdot \mathbf{E}\nonumber\\
&= \mu E_z \, \partial_z n_e + n_e E_z \, \partial_z\mu + \mu n_e\rho/\epsilon_0, 
\end{eqnarray}
where equation~\eqref{eq:model3} was used \ue{for the second line}.
Note that \ue{on the} axis $E_z$ is positive and equal to the electric field strength $E = |{\bf E}|$ for a positive streamer moving in the z-direction.

\ue{Steady propagation is assumed for several reasons:}
\begin{itemize}
  \item \jt{The PDE model can then be approximated by a set of ODEs on the axis, as shown below.}
  \item \jt{The ODE-approximation could actually apply to non-steady streamers as well, as will be tested in Section~\ref{sec:acc}.}
  \item \jt{Steadily propagating solutions of the PDE model were recently described in~\cite{Francisco2021-no1, Francisco2021-no2, Li_2022, Guo_2022}, so they are available for quantitative tests of the ODE-approximation.}
\end{itemize}

For steady streamers the temporal derivative $\partial_t$ can be replaced by a spatial derivative in the \sout{co-moving} frame 
\ue{moving with the streamer velocity $v$}:
\begin{equation} \label{eq:comoving}
\partial_t \longrightarrow - v \partial_z.
\end{equation}
Transforming to the co-moving frame (\ref{eq:comoving}) and using the identity (\ref{eq:nabla}), the model equations (\ref{eq:model1})-(\ref{eq:model2a}) become two first order ODEs (ordinary differential equations) for the electron and ion density on the axis:
\begin{eqnarray} \label{eq:model2}
    d_z n_e &=& -\,\frac{\bar{\alpha}\,\mu E\,n_e + S + E\,n_e \, d_z\mu +\mu\,{\rho \,n_e}/{\epsilon_0}}{v + \mu E},\label{eq:dz-ne-axis}\\
    d_z n_+ &=& -\,\frac{\bar{\alpha}\,\mu E\,n_e+ S}v, \label{eq:dz-ni-axis}
\end{eqnarray}
where $d_z$ denotes the derivative in the z-direction.




\subsection{An approximation for the electric field}
\label{sec:an-appr-electr}

\begin{figure}
    \centering
    \includegraphics[width=0.5\textwidth]{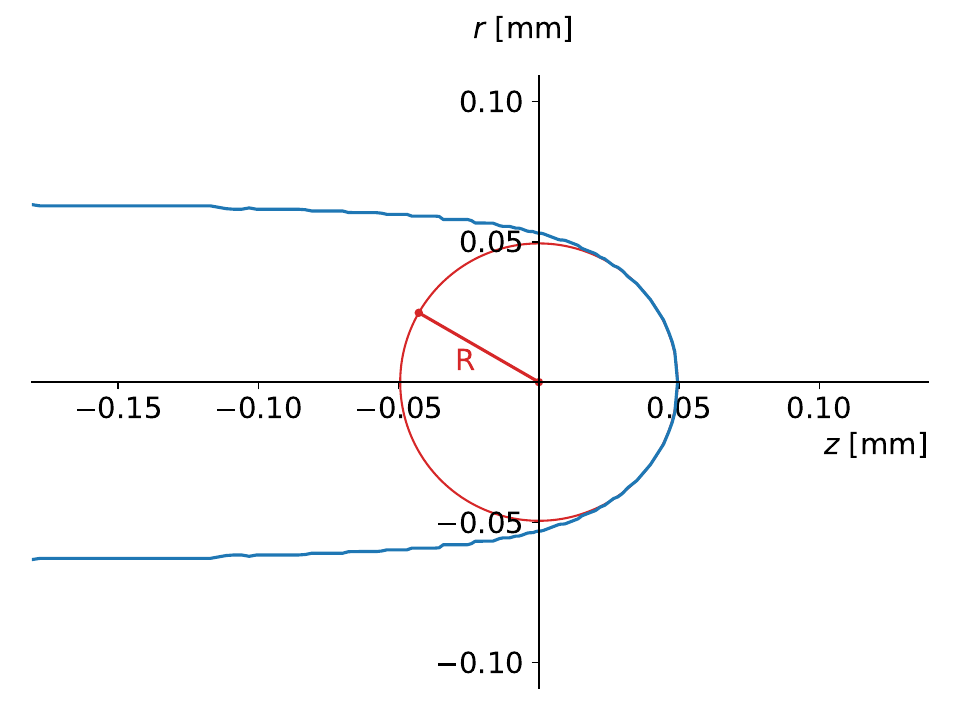}
    \caption{The charge layer of a steady streamer with photo-ionization in a background field $E_{\rm bg} = 4.5$~kV/cm according to the PDE-simulation in \cite{bouwmanEstimating2023}, in cylinder coordinates $(r,z)$ using the cylindrical symmetry around the $z$ axis. The blue line represents the position of the maximum of the charge density $\rho$ for each $z$. 
    The red line is a tangent sphere with radius $R$. At the front part, the red line is mostly hidden behind the blue line.  Note that $R$ is defined by the maximal charge density in the present figure, and by the maximal field in the rest of the paper.
    [Figure adapted from \cite{bouwmanEstimating2023}.}
    \label{fig:circle}
\end{figure}

We here present an approximation for the electric field profile on the z-axis \ue{in the streamer head region; this profile} \sout{that} can be used in equations \eqref{eq:dz-ne-axis}-\eqref{eq:dz-ni-axis}.
The basic idea is to split the field into two components
\begin{equation}
  \mathbf{E} = {\bf E}_{\rm bg} + {\bf E}_{q},
\end{equation}
where the background electric field ${\bf E}_{\rm bg}=E_{\rm bg} \hat{\bf z}$ is assumed to be homogeneous and directed in the $z$ direction, and the space charge field ${\bf E}_{q}$ is \ue{due to} \sout{assumed to be that of a (semi-)spherical charge distribution around the streamer head, as explained below.
A streamer head is surrounded by} a thin charge layer that modifies the background field.
The front part of this charge layer closely resembles a \sout{semi}sphere, as illustrated for a steady solution of the full PDE model in figure~\ref{fig:circle}.
Near the charge layer and near the axis, the resulting space charge field can therefore be approximated by that of a spherical charge distribution
\begin{equation}
  \mathbf{E}_q(r) = \frac{q(r) \, \hat{\mathbf{r}}}{4\pi \,\epsilon_0 \,r^2} \label{eq:E(R)},
\end{equation}
where $r=0$ is the center of the sphere, $\hat{\mathbf{r}}$ \ue{is a unit vector pointing} \sout{points} radially outward, and $q(r)$ denotes the charge density integrated over the volume up to radius $r$,
\begin{equation}
  q(r) = \int_0^r 4\pi r'^2\,\rho(r')\,dr', \label{eq:Int-rho}
\end{equation}
where $\rho(r')$ is a spherically symmetric charge distribution.
On the z-axis we have $r = z$, so that the z-component of $\mathbf{E} = {\bf E}_{\rm bg} + {\bf E}_{q}$ is given by
\begin{equation} \label{eq:E-approx}
    E_z(z) = E_{\rm bg} + \frac{q(z)}{4\pi\,\epsilon_0 z^2}.
\end{equation}
This approximation is valid close to the charge layer, where most ionization is produced.
It is also valid far \sout{away from} \ue{ahead of} it, where $E_z(z) \approx E_{\rm bg}$.
For $z$ similar to the streamer radius $R$, the approximation could be improved by taking the actual non-spherical charge distribution behind the streamer head into account, but we leave this for future work.

To combine equation~\eqref{eq:E-approx} with equations~\eqref{eq:dz-ne-axis}-\eqref{eq:dz-ni-axis}, 
we write equation (\ref{eq:Int-rho}) in differential form
\begin{equation}\label{eq:limit}
    d_z q(z) = 4\pi\,z^2\,\rho(z), \quad q(0)=0,
\end{equation}
where we again have used the fact that $r = z$ on axis.
This ODE can then be solved together with equations~\eqref{eq:dz-ne-axis}-\eqref{eq:dz-ni-axis}.

Finally, we remark that in the above approximation the total apparent charge $Q$ of a streamer head determines $E_q$ for any $z$ \sout{outside} \ue{ahead of} the charge layer.
There the electric field of equation \eqref{eq:E-approx} can be approximated by
\begin{equation}
  \label{eq:E-approx-far}
  E_z(z) = E_{\rm bg} + \frac{Q}{4\pi\,\epsilon_0 z^2}.
\end{equation}

\subsection{The complete ODE approximation}

On the symmetry axis the complete ODE approximation for positive streamer heads moving in the positive $z$ direction is now given by three ODEs of first order
\begin{eqnarray} \label{eq:mod1}
    d_z n_e &=& -\,\frac{\bar{\alpha}\,\mu E\,n_e + S + E\,n_e d_z\mu +\mu\,{\rho \,n_e}/{\epsilon_0}}{v + \mu E},\\
    d_z n_+ &=& -\,\frac{\bar{\alpha}\,\mu E\,n_e + S}v, \label{eq:mod2}\\
    d_z q &=& 4\pi\,z^2\,\rho, \quad \rho = e(n_+-n_e)\label{eq:mod3}
\end{eqnarray}
in an electric field 
\begin{equation} \label{eq:mod4}
    E(z) = E_{\rm bg} + \frac{q(z)}{4\pi\,\epsilon_0\,z^2}.
\end{equation}
The functions $\bar{\alpha}$, $\mu$ and $S$ depend on the local electric field, and they are specific for the respective gaseous medium.
They are an input for calculations.
The term $S$ will represent photo-ionization in section~\ref{sec:photo} and be specified there. In the cases with background ionization discussed in section~\ref{sec:Sec3}, $S$ is set to zero.

We solve the set of ODEs (\ref{eq:mod1})--(\ref{eq:mod4}) as an initial value problem, integrating \sout{backwards} from a point $z_0$ ahead of the streamer towards smaller $z$.
The following parameters have to be specified to obtain a unique solution: 
\begin{itemize}
    \item the apparent head charge $q(z_0) = Q$,
    \item the streamer velocity $v$,
    \item the background field $E_{\rm bg}$, and 
    \item the background ionization $n_e(z_0) = n_+(z_0) = n_0$ at the initial position $z_0$ ahead of the avalanche zone.
\end{itemize}

Integration proceeds through the avalanche zone and the charge layer, until the solution reaches the critical field $E_k$ in the interior (defined by $\bar{\alpha}(E_k) =0$); this point where 
\begin{equation}    \label{eq:z_ch}
    E(z_{\rm ch})=E_k,
\end{equation}
will be called $z_{\rm ch}$, where `ch' stands for channel. \ue{By construction the ion density $n_+$ is maximal at $z_{\rm ch}$; it will be denoted as $n_{\rm ch}$ below.}

Without photoionization we will \sout{use} \ue{start the integration at} $z_0 = z_k$, where $z_k$ is the location ahead of the avalanche zone where the electric field equals the critical field $E_k$,
\begin{equation}
  \label{eq:z-k}
  z_k = \sqrt{\frac Q {4 \pi \epsilon_0 (E_k - E_{\rm bg})}},
\end{equation}
according to equation~\eqref{eq:E-approx-far}.
With photoionization, we will use $z_0 = 5 z_k$, since photoionization is also produced ahead of the avalanche zone.



We numerically integrate the ODEs with an explicit finite difference scheme to obtain the complete profiles of $n_e(z)$, $n_+(z)$ and $E(z)$ on the axis.
From these profiles, we can obtain several macroscopic quantities:
\begin{itemize}
  \item The maximal electric field $E_{\rm max}$.
  \item The streamer radius $R$, which we here define as the $z$-coordinate of $E_{\rm max}$ (when the origin of the sphere is at $z=0$).
  This definition is convenient for comparison against PDE solutions.
  \item The profile of the space charge layer $\rho(z)$, from which we can also get the charge layer width $\ell$.
  \item The ion density behind the space charge layer $n_{\rm ch}$.
\end{itemize}

In summary, the ODE model with background ionization requires four input parameters: $Q$, $v$, $E_{\rm bg}$ and $n_0$, from which all macroscopic output parameters of the streamer head can be determined:
\begin{equation} \label{eq:Qv}
  (Q,\,v,\,E_{\rm bg},\,n_0) \to (R, \,E_{\rm max}, \, n_{\rm ch}, \,z_k,\,z_{\rm ch}, \,\ldots).
\end{equation}
However, the apparent charge $Q$ of the streamer head is not a parameter that can easily be measured or interpreted experimentally.
It can therefore be more intuitive to adjust $Q$ to obtain a certain radius $R$ using an optimization algorithm.
Assuming there is a one-to-one relationship between $Q$ and $R$, the input and output of the model become:
\begin{equation} \label{eq:Rv}
  (R,\,v,\,E_{\rm bg},\,n_0) \to (Q, \,E_{\rm max}, \, n_{\rm ch}, \,z_k,\,z_{\rm ch}, \,\ldots).
\end{equation}
Similarly, we can adjust $Q$ to obtain a certain $E_\mathrm{max}$ rather than an $R$.

\newpage

\section{ODE solutions with background ionization}
\label{sec:Sec3}

\db{\it Remark: Sections 3.1 and 3.2 have been exchanged, and the new Section 3.2 has been largely rewritten.}

\subsection{Results of the ODE-model with background ionization}

We now present steady streamer head solutions with background ionization, but without photoionization, so that $S = 0$ in equations \eqref{eq:mod1}--\eqref{eq:mod2}.
The integration will be performed through the active zone (i.e., avalanche zone and charge layer) where $E(z) > E_k$ and \ue{therefore} $\bar{\alpha}>0$, starting from $z_k$ given by equation~\eqref{eq:z-k}, \ue{up to $z_{\rm ch}$ of equation (\ref{eq:z_ch})}.

\begin{figure}
  \centering
  \includegraphics[width=\linewidth]{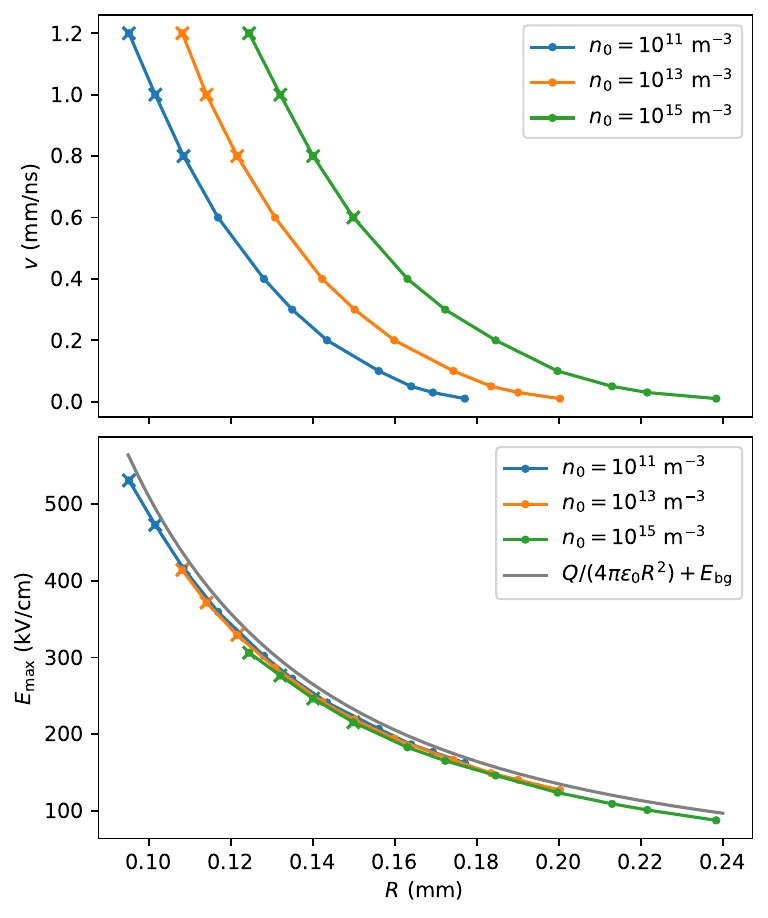}
  \caption{Solutions of the ODE model without photoionization for four parameters: $Q = 5.56 \times 10^{-11} \, \mathrm{C}$, $E_\mathrm{bg} = 10 \, \mathrm{kV/cm}$ and a range of values for the background ionization $n_0$ and velocity $v$. The corresponding radius $R$ and maximal electric field $E_{\rm max}$ are calculated. Then velocity $v$ (top) and maximal field $E_{\rm max}$ (bottom) are plotted as a function of radius $R$ and background ionization $n_0$ as indicated.  
  The crosses mark points where equation \eqref{eq:I-cond+} is not satisfied.}
  \label{fig:ode-fixed-q}
\end{figure}

We first consider a case where the solutions are identified by the four parameters $(Q,\,v,\,E_{\rm bg}, \,n_0)$, as specified by \ue{the parameter scheme} (\ref{eq:Qv}). They determine the other solution parameters like $R$ and $E_{\rm max}$. We fix $Q = 5.56 \times 10^{-11} \, \mathrm{C}$ and $E_\mathrm{bg} = 10 \, \mathrm{kV/cm}$, 
and we explore the solutions for varying streamer velocity $v$ and initial electron density $n_0$.
As for all other simulations presented in this paper, the background gas was dry air at $1 \, \textrm{bar}$ and $300 \, \textrm{K}$.
The upper plot in figure~\ref{fig:ode-fixed-q} shows $v$ and the lower plot shows $E_\mathrm{max}$ as a function of $R$ and $n_0$.

For the cases considered here, the width $\ell$ of the charge layer is much smaller than the radius $R$.
Using equation~\eqref{eq:E-approx-far}, the maximal electric field can be approximated quite well by
\begin{equation}
  \label{eq:Emax-approx}
  E_{\rm max}\approx \frac Q {4\pi\epsilon_0\,R^2} + E_\mathrm{bg}.
\end{equation}
This approximation is inserted in the lower plot in Figure~\ref{fig:ode-fixed-q}; by construction  
it fits the ODE-solutions very well.



\begin{figure}
    \centering
    \includegraphics[width=1.0\linewidth]{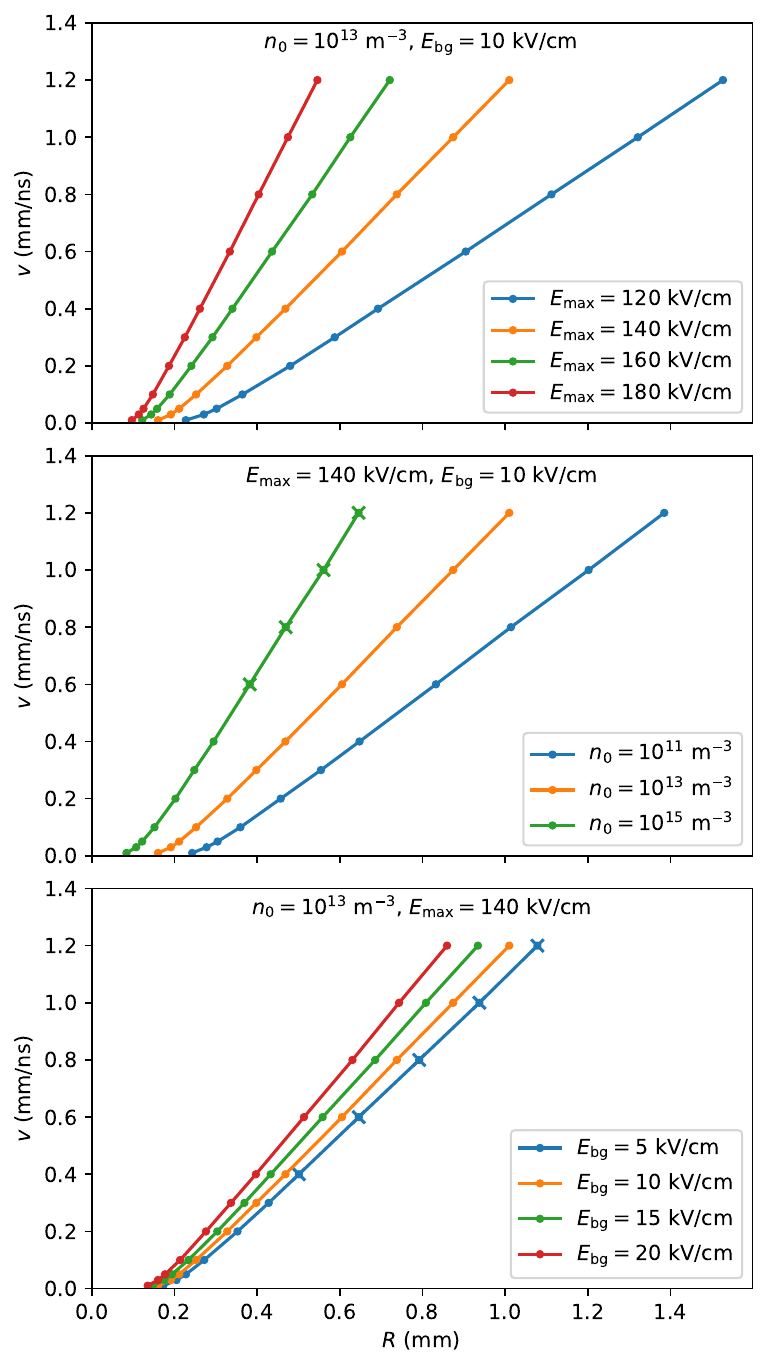}
    \caption{Velocity $v$ versus radius $R$ in ODE solutions without photoionization, for given values of $n_0$, $E_\mathrm{max}$ and $E_\mathrm{bg}$. In each figure one parameter is varied: $E_{\rm max}$ (top), $n_0$ (middle) and $E_{\rm bg}$ (bottom), while the other parameters are fixed at the indicated values.
    The crosses indicate points where equation \eqref{eq:I-cond+} is not satisfied. The numerical data of the plots are available in the supplement https://doi.org/10.5281/zenodo.14192466.}
    \label{fig:ode-fixed-Emax}
\end{figure}



Figure~\ref{fig:ode-fixed-Emax} shows the velocity $v$ versus the radius $R$ for ODE solutions in which $n_0$, $E_\mathrm{max}$ and $E_\mathrm{bg}$ were specified
\ue{according to the parameter scheme (\ref{eq:Rv})}.
Each sub-figure shows how the solutions depend on one parameter while keeping the others fixed.
A number of observations can be made:
\begin{itemize}
  \item All plots show an approximately linear relationship between velocity $v$ and radius $R$ for all parameters ($n_0$, $E_\mathrm{max}$, $E_\mathrm{bg}$). Deviations from linearity are visible for small $v$.
  \item $E_{\rm max}$ (top plot): A larger maximal electric field $E_{\rm max}$ at the same radius $R$ supports a larger streamer velocity $v$. This is due to the fact that the avalanche zone is larger and has higher fields; therefore the electron multiplication becomes faster. 
  \item $n_0$ (middle plot): For a given radius $R$ a streamer with a higher background electron density $n_0$ has a higher velocity $v$. 
  \item $E_{\rm bg}$ (bottom plot): When $E_\mathrm{max}$ and $n_0$ are kept fixed, the dependence on the background field $E_{\rm bg}$ is rather weak, since the discharge dynamics in the streamer head is dominated by the regions with high field enhancement, with fields much larger than $E_{\rm bg}$.
  Note, however, that the background field plays an important role for the streamer channel, since it determines the current in the channel (cf.\ equation~\eqref{eq:I-cond+}), and therefore the streamer head potential. 
\end{itemize}

\subsection{Constraints on steady streamer head solutions}
\label{sec:constraints}

Not all ODE-solutions for steadily propagating streamer heads will be seen in PDE-simulations or in experiments of steady streamers. At this moment, we are aware of three constraints: 
\begin{itemize}
    \item [1.] Not every ODE-solution for a streamer head is stable under the full PDE dynamics. This fact and its implications will be further discussed in section~\ref{sec:multiplicity}.
    \item[2.] The electric field integrated over the streamer length plus the additional electric potential at the streamer head have to equal the integral over the background field due to electrostatics, therefore the steady PDE solutions of a complete streamer depend on the charge distribution and hence on the conductivity over the length of the channel~\cite{bouwmanEstimating2023, starikovskiyHowPulsePolarity2020}. 
    \ue{\item[3.] If the streamer velocity is so small that it becomes comparable to the ion drift velocity in the high field region, ion motion must be included.}
    \item[4.] The electric current in the streamer has to be continuous, and the interior field in the channel \ue{typically does not} exceed the background field. \ue{As described below, this leads to the rough estimate}
    \begin{equation}
  \label{eq:I-cond+}
  \frac{\varepsilon_0}{e \mu n_\mathrm{ch}} \, \frac{v}{R} \, \left (\frac{E_\mathrm{max}}{E_\mathrm{bg}} - 1 \right ) < c_0,
\end{equation}
\jt{where $c_0$ is a constant less than one, but of order unity.}
\end{itemize}

\sout{We now study what the last constraint implies for the ODE solutions of the streamer heads,} \ue{To derive (\ref{eq:I-cond+}) we evaluate the continuity of the electric current behind the space charge layer. On the streamer axis at the center of the sphere, this current can be estimated as}
\begin{equation}
  \label{eq:I-ch+}
  I_\mathrm{ch} \approx \pi R^2 e \mu E_\mathrm{ch} n_\mathrm{ch},
\end{equation}
where $\pi R^2$ is the channel's cross-sectional area, $E_\mathrm{ch}$ is the electric field in the channel, and $e \mu E_\mathrm{ch} n_\mathrm{ch}$ is the conduction current density. 

\ue{On the other hand, the current $I_{\rm head}$ in the streamer head can be roughly estimated from the apparent head charge $Q$, radius $R$ and velocity $v$ of the streamer if the streamer moves steadily and if there are no currents ahead of it, as recalled in \cite{Guo_2022}. We use that the line charge } 
\begin{equation}
        \lambda(z)=\int_0^\infty 2\pi r \,\rho(r,z) \,dr
\end{equation}
\ue{of a semi-sphere with homogeneous surface charge and total charge $Q$ is constant and equal to $Q/R$ on the axis.} And if a line charge moves with velocity $v$, it carries a current 
\begin{equation}
  \label{eq:I-head+}
  I_\mathrm{head}(z) = v\;\lambda(z)\approx v\;Q/R. 
\end{equation}
\ue{Here the real charge distribution is approximated by the apparent head charge $Q$.}

  
\jt{Using equations~\eqref{eq:I-ch+} and \eqref{eq:I-head+} }
\ue{and expressing $Q$ by $4 \pi \varepsilon_0 R^2 (E_\mathrm{max} - E_\mathrm{bg})$ according to Equation~\eqref{eq:E-approx-far}, the continuity} 
condition $I_\mathrm{ch} = I_\mathrm{head}$ can be written as
\begin{equation}
  \label{eq:I-equal+}
  \frac{\varepsilon_0}{e \mu n_\mathrm{ch}} \, \frac{v}{R} \, \frac{E_\mathrm{max} - E_\mathrm{bg}}{E_\mathrm{ch}} \approx 1/4.
\end{equation}
We cannot compute $E_\mathrm{ch}$ in our ODE model, but \jt{if we assume that the channel field $E_{\rm ch}$ is smaller than the background field $E_{\rm bg}$}, equation~\eqref{eq:I-equal+} implies \ue{the final result} (\ref{eq:I-cond+}) with $c_0 \sim 1/4$.

\ue{The parameter sets in figures~\ref{fig:ode-fixed-q} and \ref{fig:ode-fixed-Emax} that do not satisfy equation~\eqref{eq:I-cond+} with $c_0 = 1/4$ are marked with crosses. They do not allow a continuous current and line charge in the streamer head, if the channel field $E_{\rm ch}$ is below the background field $E_{\rm bg}$.}

\newpage

\section{ODE and PDE solutions with photoionization}
\label{sec:photo}

Instead of a background ionization $n_0$, we will now include photoionization as a source of free electrons ahead of a positive streamer.
Since photoionization is typically the dominant source of free electrons ahead of streamers in air, we will neglect background ionization when photoionization is included, assuming $n_0 = 0$.

\subsection{Including photoionization in the ODE model}
\label{sec:incl-phot-ode}

We use Zheleznyak's classical model~\cite{zheleznyak_photoionization_1982} for air, in which the photoionization source term $S_{\rm ph}$ is given by
\begin{equation}\label{eq:Sph}
  S_{\rm ph}(\mathbf{r}) = \int\frac{I_\mathrm{ph}(\mathbf{r}')f(|\mathbf{r}-\mathbf{r}'|)}{4\pi|\mathbf{r}-\mathbf{r}'|^2}d^3r',
\end{equation}
where the absorption function $f(r)$ is
\begin{equation} \label{eq:30}
    f(r) = \frac{\exp(-\chi_{\rm min} p_{O_2}r) - \exp(-\chi_{\rm max} p_{O_2}r)}{r\ln(\chi_{\rm max}/\chi_{\rm min})},
\end{equation}
with $\chi_{\rm max} = 150$ (mm bar)$^{-1}$, $\chi_{\rm min} = 2.6$ (mm bar)$^{-1}$ and $p_{O_2}$ the partial pressure of oxygen. 
Furthermore, the source of ionizing photons $I_\mathrm{ph}(\mathbf{r})$ is given by
\begin{equation}
    I_\mathrm{ph} = \frac{p_q}{p+p_q}\xi S_i,
\end{equation}
where $S_i$ is the electron impact ionization source term, $p_q=40$ mbar is the quenching pressure, and $\xi$ is the proportionality factor between impact ionization rate and photon emission rate, that we will approximate by a constant $\xi=0.075$, as in our previous work~\cite{li2021, wang2023}.

As discussed and approximated in~\cite{bouwmanEstimating2023}, most ionizing photons are produced close to the charge layer, since most of the ionization reactions occur there.
However, in our present ODE model, we will for simplicity assume that photons are produced by a point source, which avoids a multidimensional integral.
This is a \jt{reasonable} approximation when the streamer radius is smaller than the typical absorption length of photons \ue{that reaches here up to 2 ~mm}.
Since the charge layer corresponds approximately to a semi-sphere between $z = 0$ and $z = R$, see section \ref{sec:an-appr-electr}, we here place the point source at $z = R/2$.
\jt{We remark that there is actually a distribution of absorption lengths, see equation (\ref{eq:30}), and the point-source approximation will be less accurate for shorter absorption lengths, while the longer lengths contribute more to the motion.}

The total number $I_{\rm tot}$ of ionizing photons produced per unit time and integrated over space, is given by
\begin{equation}
  \label{eq:I-tot}
    I_{\rm tot} = \frac{p_q}{p+p_q}\xi S_\mathrm{tot} = \frac{p_q}{p+p_q}\xi \pi R^2 n_\mathrm{ch} v,
\end{equation}
where we have approximated the total amount of impact ionization per unit time by $S_\mathrm{tot} = \pi R^2 n_\mathrm{ch} v$, with $v$ being the streamer velocity and $n_{\rm ch}$ the electron density in the channel. 
With the above approximations, photoionization in the ODE model is given by
\begin{equation}\label{eq:Sph-ode}
  S_{\rm ph}(z) = I_{\rm tot} \,\frac{f(z-R/2)}{4\pi(z-R/2)^2}.
\end{equation}

Solving the ODE model becomes an implicit problem when photoionization is considered, since the photoionization source term depends on $I_{\rm tot}$ and $R$, which are not known beforehand.
Therefore we adopt an iterative approach to find $I_{\rm tot}$ and $R$, by using a root-finding algorithm on the following function
\begin{equation}
  \label{eq:root-eq}
  g(I_{\rm tot}^*, R^*, Q, v, E_\mathrm{bg}) = (I_{\rm tot}^* - \tilde{I}_{\rm tot}, R^* - \tilde{R}),
\end{equation}
where $I_{\rm tot}^*$ and $R^*$ are the current guesses for $I_{\rm tot}$ and $R$.
The right-hand side vector contains $\tilde{R}$ and $\tilde{I}_{\rm tot}$, which are obtained by first solving the ODE model using the parameters $(I_{\rm tot}^*, R^*, Q, v, E_\mathrm{bg})$, and then evaluating the resulting radius and total amount of ionizing photons~\eqref{eq:I-tot}.
Note that the next guesses for the parameters $I_{\rm tot}^*$ and $R^*$ will be determined by the root-finding algorithm until it is converged.
Assuming that background ionization is negligible compared to photoionization (i.e., $n_0 = 0$), the ODE model requires now only three input parameters 
\begin{equation} \label{eq:vRphotoQ}
    (Q,\, v,\, E_\mathrm{bg}) \to (R, \,E_{\rm max}, \, n_{\rm ch}, \,z_k,\,z_{\rm ch}, \,\ldots)
\end{equation}
to determine a steady head solution.
As before, we can replace the input parameter $Q$ by $E_\mathrm{max}$ or $R$, by searching for the value of $Q$ that corresponds to a particular $E_\mathrm{max}$ or $R$. In particular, we get 
\begin{equation} \label{eq:vRphotoR}
    (R,\, v,\, E_\mathrm{bg}) \to (Q, \,E_{\rm max}, \, n_{\rm ch}, \,z_k,\,z_{\rm ch}, \,\ldots).
\end{equation}

The python code used to implement this calculation is available in the supplement \\ https://doi.org/10.5281/zenodo.14192466.


\begin{figure}
  \centering
  \includegraphics[width=\linewidth]{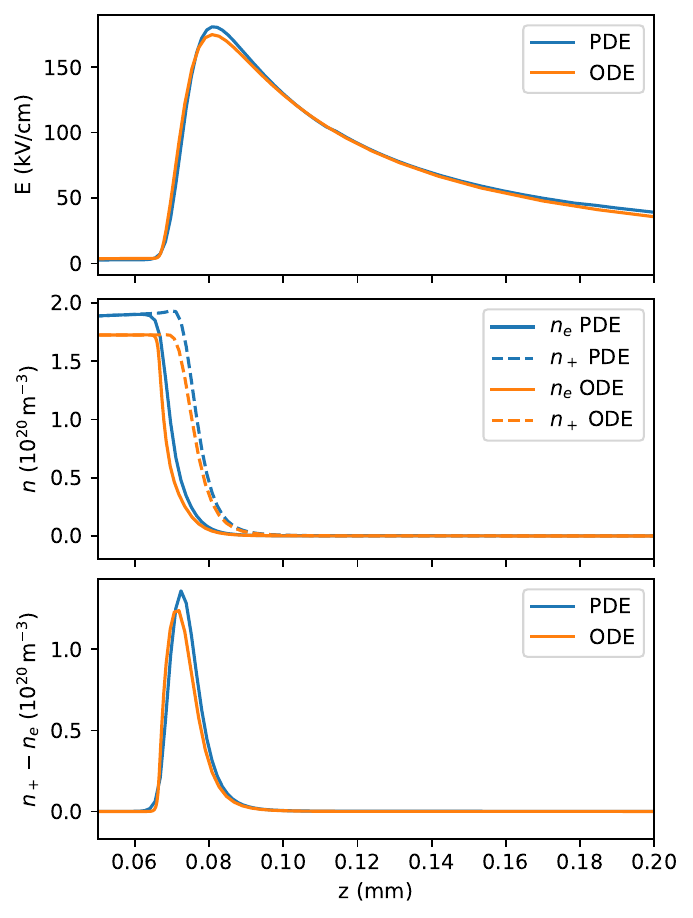}
  \caption{Comparison of \ue{the spatial profiles of electric field and densities of} the ODE model against steady positive streamers simulated with a PDE model~\cite{Li_2022}. Both types of simulations were performed in air with photoionization. The ODE solution was computed by specifying three parameters from the PDE solution: $v = 0.08 \, \textrm{mm/ns}$, $R = 80.7 \, \mu\mathrm{m}$ and $E_\mathrm{bg} = 4.33 \, \mathrm{kV/cm}$. The PDE solution was shifted to align the maxima in the electric field.}
  \label{fig:comparison_v8_xiaoran}
\end{figure}

\subsection{Definition of the radius}
\label{sec:radius}

We will now compare solutions of the ODE model with full PDE solutions.
Since it is important to use a consistent definition of the streamer radius $R$ in such a comparison,
we will fit the on-axis electric field profile in the PDE solution with a quadratic decay of the form
\begin{equation}
  \label{eq:radius-fit}
  E_\mathrm{fit}(z) = E_\mathrm{bg} + (E_\mathrm{max} - E_\mathrm{bg})
  \left(\frac R {R + z - z_\mathrm{head}}\right)^2,
\end{equation}
where $z_\mathrm{head}$ is the $z$-coordinate at which the maximum electric field $E_\mathrm{max}$ occurs.
The only unknown in this expression is $R$, which we obtain by fitting between the location where $|\partial_z E(z)|$ has a maximum (which lies just outside the streamer's charge layer) and the location where $E(z) = E_\mathrm{max}/2$.
Equation~\eqref{eq:radius-fit} is consistent with the definition of the radius that we use in the ODE model, which can be seen by combining equations~\eqref{eq:E-approx-far} and \eqref{eq:Emax-approx}.

\subsection{Comparison of steady PDE and ODE solutions}
\label{sec:res_Sph}

To compare the intrinsically steady ODE solutions with PDE solutions, steady PDE solutions are required. Note that there is an old assumption that only one background field, called the stability field~\cite{Li_2022}, would support steadily propagating streamers. But we recently have found that there is a range of background electric fields that support steady propagation of streamers with different radii and velocities. Steady streamer solutions of the PDE model 
exist in background fields ranging from $4.0$ to $5.5$ kV/cm \cite{Li_2022}; and each solution has a specific velocity in the range from $0.03 \, \textrm{mm/ns}$ to $0.12 \, \textrm{mm/ns}$.
In figure~\ref{fig:comparison_v8_xiaoran} we compare our ODE model with the PDE solution from~\cite{Li_2022} that corresponds to $v = 0.08 \, \textrm{mm/ns}$, $R = 80.7 \, \mu\mathrm{m}$ and $E_\mathrm{bg} = 4.33 \, \mathrm{kV/cm}$.
(Note that this value of $R$ differs from the one given in table B1 of~\cite{Li_2022}, since we here use the radius definition~\eqref{eq:radius-fit}.)
The resulting profiles of the electric field $E$, electron density $n_e$, positive ion density $n_+$, as well as of the charge density $e (n_+ - n_e$) presented in Figure~\ref{fig:comparison_v8_xiaoran} all show good agreement.
In the ODE model $E_\mathrm{max}$ is slightly lower, which leads to a lower electron density $n_e$ in the channel and in the charge layer, according to an old estimate that was recently re-analyzed and improved in~\cite{bouwmanEstimating2023}.

Besides the obvious difference in dimensionality and in how the electric field is computed, there are a number of other differences between the ODE and PDE models:
\begin{itemize}
  \item Electron diffusion is not included in the ODE model, whereas the PDE model included diffusion plus a correction of the source term to avoid unphysical effects due to diffusion (see~\cite{Li_2022} for details).
  \item The PDE simulations included different types of ions and electron detachment, whereas all ion species are grouped into a single species $n_+$ in the ODE model. 
  \item The PDE simulations included ion drift.
  \item Photoionization is approximated by a point source at the location $z=R/2$ in the ODE model, as described in section~\ref{sec:incl-phot-ode}, whereas the so-called Helmholtz approximation was used in the PDE simulations.
\end{itemize}
Despite these differences, the solutions shown in figure~\ref{fig:comparison_v8_xiaoran} agree very well, which indicates that the approximations made in the ODE model are valid under these conditions.

\begin{figure}
    \centering
    \includegraphics[width=\linewidth]{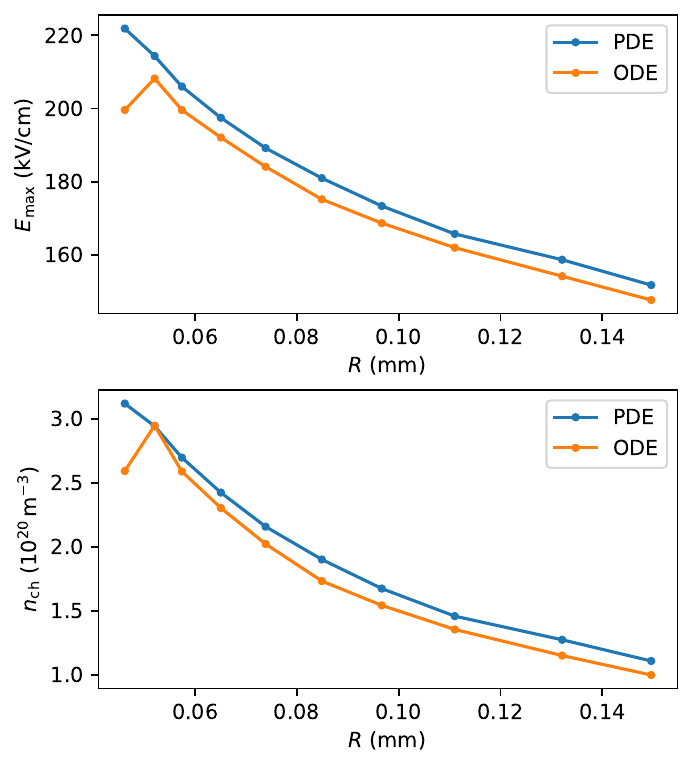}
    \caption{Comparison of the steady PDE solutions with photoionization from~\cite{Li_2022} 
    with the corresponding ODE solutions. The steady PDE solutions are uniquely determined by their radius $R$, and their maximal field and ionization density in the channel are plotted as a function of $R$. 
    The parameters $v$, $R$ and $E_\mathrm{bg}$ of the PDE solutions were identified and used to compute the ODE solutions that are shown as well.}
    \label{fig:comparison_sweep_photoi_xiaoran}
  \end{figure}

In figure~\ref{fig:comparison_sweep_photoi_xiaoran} we compare the steady ODE and PDE solutions for all steady streamers from~\cite{Li_2022} and for the corresponding ODE solutions. For each steady PDE solution the parameters $v$, $R$ and $E_\mathrm{bg}$ were identified and the corresponding ODE solution
calculated. Figure~\ref{fig:comparison_sweep_photoi_xiaoran} shows the maximal field $E_\mathrm{max}$ and the channel density $n_\mathrm{ch}$ as a function of streamer radius $R$. 
Except for the leftmost data point, the curves generally show good agreement between PDE and ODE solutions with differences being up to 3\% in $E_\mathrm{max}$ and up to 10\% in $n_\mathrm{ch}$.
The leftmost data point corresponds to the case with the smallest radius and smallest velocity ($0.03 \, \textrm{mm/ns}$), and there differences are about 10\% and 20\% respectively.

\subsection{Relation between velocity, radius and maximal field}
\label{sec:Emax}

\begin{figure}
  \centering
  \includegraphics[width=\linewidth]
  {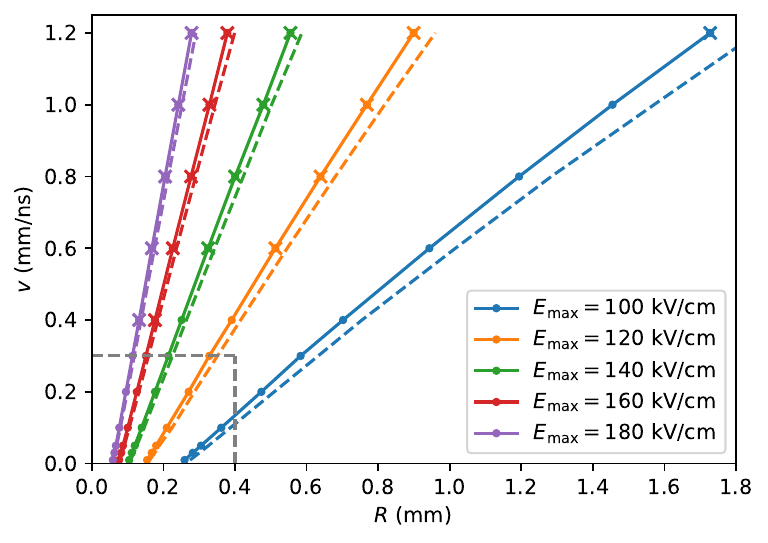}
  \includegraphics[width=\linewidth]
  {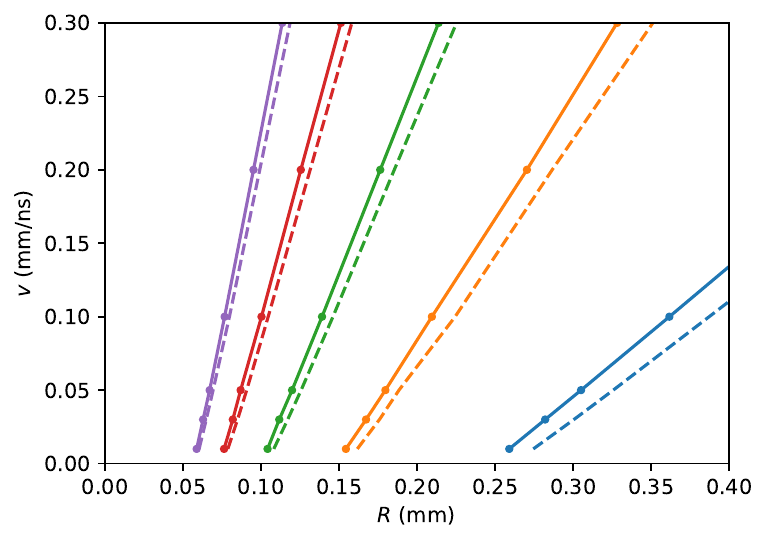}
  \caption{The upper plot shows the maximal field $E_{\rm max}$ as a function of velocity $v$ and radius $R$ for streamers in dry air (1 bar, 300 K) with photoionization, computed with the ODE model in a fixed background field $E_{\rm bg}=10$ kV/cm (solid lines) and 5~kV/cm (dashed lines).
    \jt{The lower plot zooms into the dashed gray box in the upper plot.}
    The plots can be used to derive the maximal electric field from radius and velocity of a steadily propagating streamer. The numerical data of the plot are available in the supplement https://doi.org/10.5281/zenodo.14192466.}
  \label{fig:vR-relation-Ebg}
\end{figure}

In (\ref{eq:vRphotoR}) we already have noted that the three parameters of velocity $v$, radius $R$ and background field $E_{\rm bg}$ determine a unique solution for steady streamer heads with photoionization. 
Figure \ref{fig:vR-relation-Ebg} shows 
\begin{equation}
    E_{\rm max}= E_{\rm max}(v,R,E_{\rm bg})    
\end{equation}
as a function of $v$ and $R$ in a background field of $E_{\rm bg}=10$~kV/cm as solid lines, and for $E_{\rm bg}=5$~kV/cm as dashed lines. \ue{The lower plot zooms into the upper plot, showing the data for small $v$ and $R$.} 
Since the dependence on $E_{\rm bg}$ is quite weak, we can to a good approximation describe a steadily propagating streamer head in air (with photoionization) by just its velocity and radius, and the maximal field is then parameterized as $E_{\rm max} \approx E_{\rm max}(v,R)$; this allows to determine the maximal field from velocity and radius.

The apparent head charge is approximated by 
\begin{equation}\label{eq:Q-approx}
    Q\approx 4\pi\epsilon_0\,R^2\, (E_{\rm max}-E_{\rm bg})
\end{equation}
according to equation~(\ref{eq:Emax-approx}).

A similar analysis, but for streamers with background ionization $n_0$ was already presented in Figure~\ref{fig:ode-fixed-Emax}.
The dependence on the background field $E_{\rm bg}$ was also quite weak there, but when the background ionization level $n_0$ was increased by two orders magnitude, the streamer velocity $v$ for a given $E_\mathrm{max}$ and $R$ increased substantially.
For the results presented in figure~\ref{fig:vR-relation-Ebg} 
\ue{we expect to find} \sout{there will be} a similar dependence on the amount of photoionization.

\ue{It is interesting to note that there is an approximately linear relation between velocity $v$ and radius $R$ for fixed $E_{\rm max}$, both with background ionization in Fig.~\ref{fig:ode-fixed-Emax} and with photoionization in Fig.~\ref{fig:vR-relation-Ebg}, but $v$ and $R$ are not proportional since the curves do not go through the origin.}




Finally Naidis has suggested a relation between velocity, radius and maximal field in~\cite{naidisPositiveNegativeStreamers2009}, and we have included a discussion and comparison in \ref{app:discussion_Naidis}.
Naidis' approach differs from ours in several ways: he derived his electric field profile from a fit to simulations \ue{rather than from a consistent calculation of electron and charge densities coupled to the electric field}, and he assumed that the charge layer is located where the electron density has multiplied by a factor of $e^8$ while \ue{other dependencies are not analyzed.} Nevertheless his more heuristic result agrees with ours \ue{in Figure~\ref{fig:vR-relation-Ebg}} quite well \ue{for air under normal conditions}.


\subsection{Accelerating streamers with photoionization}
\label{sec:acc}

\begin{figure}
  \centering
  \includegraphics[width=\linewidth]{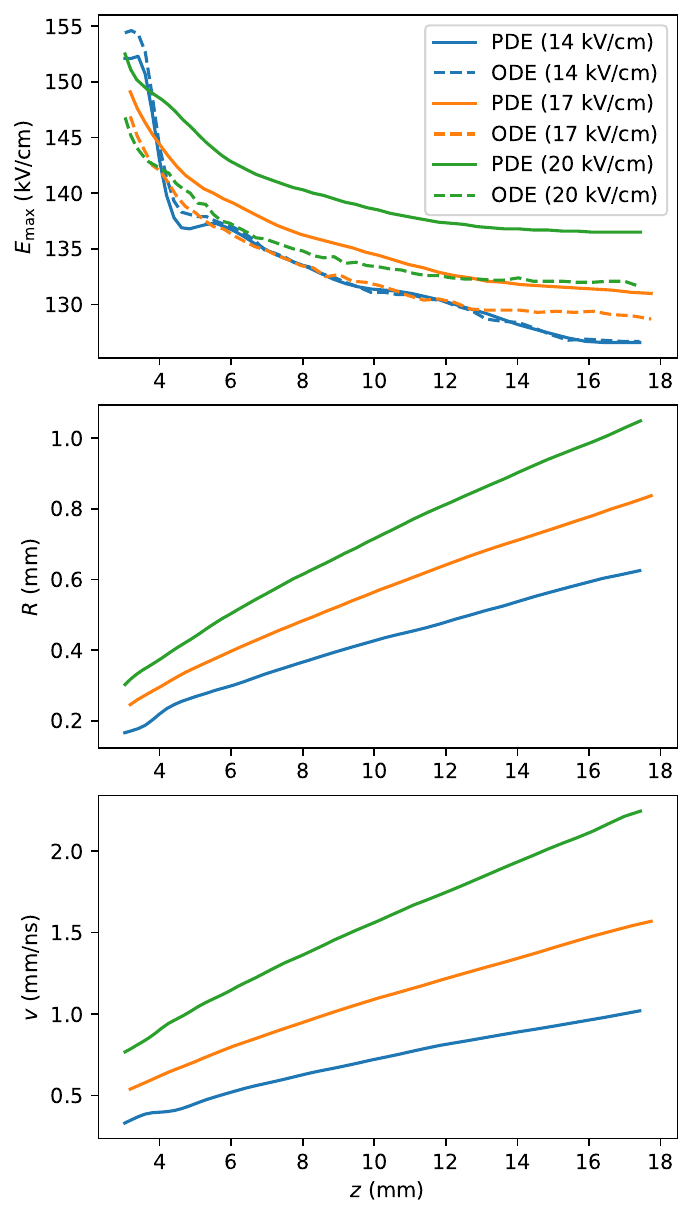}
  \caption{Comparison of the ODE model against PDE solutions of accelerating streamers. \ue{The coordinate $z$ parameterizes the location of the streamer heads while they propagate.} The simulations were performed in air with photoionization, in three different background fields. The solid lines show how $E_\mathrm{max}$, $R$ and $v$ evolve in the PDE simulations. The dashed lines \ue{in the upper plot} show the ODE solution for $E_\mathrm{max}$ using $R$, $v$ and the background electric field as input.}
  \label{fig:comparison_sweep_photoi}
\end{figure}

The ODE model was derived for steady streamers, so it is interesting to test how well it can predict the properties of accelerating streamers.
We therefore have simulated accelerating streamers with an axisymmetric PDE model as described in~\cite{teunissen_simulating_2017}.
The simulations were performed in air (at 1 bar, 300 K), in a computational domain measuring $32 \, \textrm{mm}$ in the $r$ and $z$ direction.
The domain had a plate-plate electrode geometry.
To provide initial electric field enhancement, a rod electrode with a semi-spherical cap was included, which was $1.8 \, \textrm{mm}$ long and had a radius of $0.25 \, \textrm{mm}$.
Furthermore, a low level of background ionization of $n_0 = 10^{10} \, \textrm{m}^{-3}$ was included so that a discharge could start, after which this density became negligible compared to the ionization produced by photoionization.
Three different background electric fields $E_\mathrm{bg}$ were applied: $14 \, \textrm{kV/cm}$, $17 \, \textrm{kV/cm}$ and $20 \, \textrm{kV/cm}$.

From the PDE simulations, we extracted maximal electric field $E_\mathrm{max}$, streamer radius $R$, and velocity $v$ as a function of the streamer head position $z$, as shown as solid lines in the three panels of figure~\ref{fig:comparison_sweep_photoi}.
Using $R(z)$, $v(z)$ and $E_\mathrm{bg}$ at given head positions $z$ as input, we have computed ODE solutions corresponding to these accelerating streamers, and the resulting values of $E_\mathrm{max}(z)$ are shown as dashed curves in the upper plot in figure~\ref{fig:comparison_sweep_photoi}.
In the lowest background field ($E_\mathrm{bg} = 14 \, \textrm{kV/cm}$) there is good agreement between the values of $E_\mathrm{max}(z)$ during the streamer evolution in the PDE and ODE models.
This means that the degree of ionization in the channel will also agree well, since this property is mostly determined by $E_\mathrm{max}$.
When the background field is increased to $17 \, \textrm{kV/cm}$ and $20 \, \textrm{kV/cm}$, the differences in $E_\mathrm{max}$ increase to about 2\% and 4\% respectively, with $E_\mathrm{max}$ being lower in the ODE model.
The relatively good agreement obtained here indicates that the ODE approximation retains most of its validity when studying accelerating streamers.

\newpage

\section{Summary and outlook}

\subsection{Summary}

We have developed and studied macroscopic parameterizations of streamer heads that appear to be soliton-like \sout{self-stabilized} nonlinear structures. Our starting point was the classical fluid model for streamers. We had two main motivations:
\begin{itemize}
    \item[1.] To construct dynamical building blocks (like heads and channels) for efficient multi-streamer models, and
    \item[2.] to evaluate parameters such as maximal electric field and chemical activation in the streamer channel from experimental measurements of $v$ and $R$ through the approximation 
    \begin{equation} \label{eq:Emax-sum}
    E_{\rm max} = E_{\rm max}(v,R,E_{\rm bg}).
\end{equation}
\end{itemize}
In the present paper we have studied positive streamer heads in air (at 1~bar and 300~K) that are propagating steadily, and we have found that our macroscopic ODE-approximation fits the full PDE solutions of the classical fluid approximation well. First tests show promising agreement for accelerating streamers as well. Our ODE streamer head model is defined on the streamer axis, but takes the 3D structure of the head into account through a \sout{(semi-)}spherical approximation for the charge distribution around the \ue{front part of the} streamer head and through the corresponding electric field profile. \sout{The model} 
\ue{This ODE-approximation} therefore can be seen as a 1.5D model. (The emphasis in our previous paper~\cite{bouwmanEstimating2023} was on the whole streamer and on heuristic approximations that would allow analytical approximations, while we here work with the full equations on the axis of the streamer head, and we solve them numerically.)

We have started with streamers with background ionization $n_0$ (that occurs, e.g., by repetitive pulsing or by radiation), but without photoionization. The ODE-approximation then takes the form of an initial value problem for three coupled ODEs, and a unique solution is determined by the two external parameters of background field $E_{\rm bg}$ and background ionization $n_0$ and by the two internal parameters of velocity $v$ and radius $R$. These four parameters determine the full profile of electron and ion density and electric field on the axis of a steady streamer head, 
and therefore also the macroscopic parameters of maximal field $E_{\rm max}$, apparent head charge $Q$, width $\ell$ of the charge layer, and ionization $n_{\rm ch}$ in the channel. 

Streamers with photoionization require an iterative approach, and only three parameters need to be fixed, as background ionization $n_0$ typically is dominated by photoionization. 
In particular, the maximal electric field $E_{\rm max}(v,R,E_{\rm bg})$ (\ref{eq:Emax-sum})
depends on $v$, $R$ and on the background field $E_{\rm bg}$.  But the dependence on $E_{\rm bg}$ is quite weak because the high field zone $E>E_k$ dominates the ionization reactions. 
The central result (\ref{eq:Emax-sum}) is evaluated and quantified in figure~\ref{fig:vR-relation-Ebg}. The velocity-radius relation (\ref{eq:Emax-sum}) revises an older approximation by Naidis~\cite{naidisPositiveNegativeStreamers2009} that is discussed in the appendix, and it provides an experimental access to $E_{\rm max}$ by measuring $v$ and $R$, see our second motivation. $E_{\rm max}$ in turn determines the ionization $n_{\rm ch}$ and plasma-chemical activation of the streamer channel; here we refer to a revisited  approximation of $n_{\rm ch}(E_{\rm max},\ell,R)$ in~\cite{bouwmanEstimating2023}.

Furthermore, if photoionization is absent, the background ionization $n_0$ has a strong influence on the streamer head solutions, see figure~\ref{fig:ode-fixed-Emax}. While the background field $E_{\rm bg}$ influences the head solutions only weakly, it is relevant for the currents in the streamer channel and therefore for the head potential \cite{bouwmanEstimating2023}.

\subsection{Multiplicity of solutions and dynamical selection} \label{sec:multiplicity}

We now discuss the multiplicity of steady streamers with photoionization. 

That complete streamers consisting of head and channel can propagate steadily in the full PDE fluid model was found in~\cite{Qin_2014a,Francisco2021-no1, Francisco2021-no2}. The PDE solutions of \cite{Li_2022,Guo_2022} showed that for a given background field within a certain range, there is one unique steady streamer solution (up to translation), and no free parameter. On the other hand, we here found that within a given background field there is a two-parameter family of ODE solutions for steady streamer heads, parameterized by $v$ and $R$. 

It is encouraging that the ODE head solutions reproduce the full PDE streamer solutions well, when the same parameters $v$, $R$ and $E_{\rm bg}$ are chosen, see figures~\ref{fig:comparison_v8_xiaoran} and \ref{fig:comparison_sweep_photoi_xiaoran}. But it is surprising at first sight that for given $E_{\rm bg}$ the family of ODE head solutions is characterized by two parameters $v$ and $R$, while the whole streamer PDE solution has no free parameter at all (up to translation). One would expect that connecting a streamer head with a channel would deliver one constraint on the head model, hence there should be a one-parameter family of PDE solutions for the complete streamer. 

The likely solution is that not all ODE head solutions are dynamically stable under the full PDE dynamics. This is a common feature in pattern formation and has already been demonstrated for streamers. In particular, a periodic pattern of identical and parallel negative streamers without photoionization in a 2D cartesian system in a constant electric field has been investigated in~\cite{luqueSaffmanTaylorStreamersMutual2008}. When $d$ is the streamer diameter and $D$ the length of the period, steady solutions exist for any value of $d/D$ between 0 and 1, but the streamers always approach the solution with $d/D=1/2$ in time under the full PDE dynamics, as is demonstrated in~\cite{luqueSaffmanTaylorStreamersMutual2008}. So these particular streamer solutions are called dynamically selected. We hypothesize that here as well the full PDE dynamics admits only a one-parameter family of solutions of dynamically stable streamer heads, but this question of dynamical selection is left to future investigations. 

A practical consequence is, e.g., that for each measured $v$, $R$ and $E_{\rm bg}$ the maximal field $E_{\rm max}$ of a steady solution can be read from figure~\ref{fig:vR-relation-Ebg}. But most likely not every ODE solution parameterized by $v$, $R$ and $E_{\rm bg}$ is dynamically stable. For given $E_{\rm bg}$ only a particular combination of $v$ and $R$ might yield a dynamically stable solution. Or said differently: for every radius $R$ and background field $E_{\rm bg}$, there are steady solutions for a range of velocities $v$, each one determining a different value of $E_{\rm max}$. 
The meanwhile historical question of ``What defines the radius and the maximal electric field of the streamer head?'' (posed, e.g., in the supplement of \cite{Bazelyan}) requires the solution of this dynamical selection problem.

\subsection{Outlook}

The present analysis is but a first step towards an efficient multi-streamer model along the lines laid out in~\cite{luqueGrowingDischargeTrees2014}.
Challenges already mentioned above are multiplicity and dynamic selection of the ODE-head solutions. Which ones of these head solutions are dynamically stable under the full PDEs? 
Also whether the ODE-approximations can be applied to accelerating or decelerating streamers requires more  investigations. 


On the way to multi-streamer models there are at least two more challenges: How to characterize and include streamer branching?
And how to match the streamer head model to a streamer channel model as presented by Luque~\cite{luqueStreamerDischargesAdvancing2017}?
And clearly negative streamers in air and streamers in other gases should be studied as well. 
\\

{\bf Data availability statement}

Input data, python code for the ODE-model and the numerical data presented in Figures 3 and 6 are openly available at the following URL/DOI: https://doi.org/10.5281/zenodo.14192466 .

\newpage

\appendix

\section{Approximations for $E_{\rm max}(v,R,E_{\rm bg})$ by Naidis and in the present paper 
}\label{app:discussion_Naidis}

In his much quoted paper \cite{naidisPositiveNegativeStreamers2009} on the velocity radius relation of streamers in air, 
Naidis has suggested a relation between velocity $v$, radius $R$ and maximal field $E_{\rm max}$ as well. Here we compare his approach with ours.

Naidis neglects electron diffusion as we do, arguing with the smooth decay of the front due to background ionization or photoionization. He neglects photoionization and also the background field, whereas we include \ue{both}. \sout{an $E_{\rm bg}$.} 
He then studies steady streamers in a frame co-moving with velocity $v$ as we do, but in 1D approximation. His equation (3) is in the notation of the present paper 
\begin{equation} \label{eq:A1}
     -vd_z n_e -d_z (\mu E\,n_e) = \bar{\alpha}\,\mu E\,n_e.
\end{equation}
In this 1D approximation, he therefore replaces $\nabla \cdot {\bf E}$ by $d_z E_z$ within the term $d_z (\mu E\,n_e)$; this is problematic in the charge layer.
He then integrates equation (\ref{eq:A1}) to his equation (4) 
\begin{eqnarray} \label{eq:A2}
     &&n_e(z) \,[v+\mu E(z)] \\
     &&= n_e(z_0)\, [v+\mu E(z_0)]\,
     \exp\int_z^{z_0} \frac{\bar{\alpha} \,\mu E\,dz}{v+\mu E},
     \nonumber
\end{eqnarray}
valid for any $z$ and $z_0$ on the streamer axis within the avalanche zone. 

The only remainder of the curvature of the streamer head in his 1D approach is the assumption that the electric field $E(z)$ decays in the avalanche zone. This function $E(z)$ is fitted to earlier results of a fluid PDE simulation model with the ansatz $E(z)=a/(z-z_1)$, rather than calculated self-consistently with the ansatz of a constant background field plus a term decaying quadratically like $b/(z-z_2)^2$ as in equation~(\ref{eq:E-approx}). Naidis uses his $E(z)$ to integrate equation~(\ref{eq:A2}). 

That the electric field lines converge towards the streamer tip and that therefore a larger electron density drifts towards the tip, is a 3D effect that is not included \ue{in \cite{naidisPositiveNegativeStreamers2009}}.

Furthermore, Naidis does not calculate the position of the charge layer self-consistently, but he uses the ad hoc criterion that the electron density increases by a factor $e^8$ from the front edge $z_k$ of the active region to the position $z_{\rm ch}$ behind the charge layer:
\begin{equation}  \label{eq:Naidis}
    n_{\rm ch}/n_k=e^8.
\end{equation}
In contrast, in our ODE approximation scheme, we calculate the profiles of the electric field, of the electron current density and of the charge layer self-consistently.
We find that $E_{\rm max}$ depends not only on $v$ and $R$, but also weakly on the background field $E_{\rm bg}$ and either on background ionization $n_0$ as shown in Figure~\ref{fig:ode-fixed-Emax} or on photoionization as shown in Figure~\ref{fig:vR-relation-Ebg}.

\begin{figure}
  \centering
  \includegraphics[width=\linewidth]{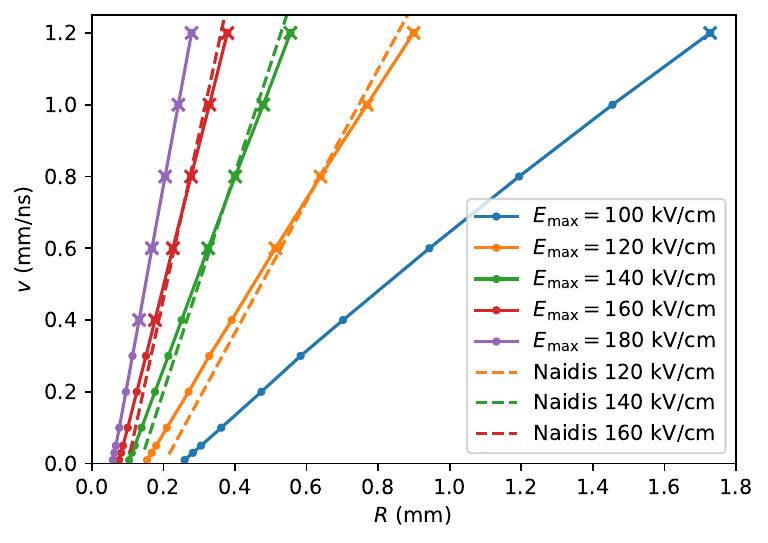}
  \caption{The maximal field $E_{\rm max}$ as a function of velocity $v$ and radius $R$ for streamers in dry air (1 bar, 300 K).
  Solid lines: the results of the ODE model with photoionization in a fixed background field $E_{\rm bg}=10$ kV/cm, reproduced from figure~\ref{fig:vR-relation-Ebg}. Dashed lines: Naidis' 2009 approximation. Note that Naidis used the ``radiation diameter'', which we have here simply divided by two to obtain a radius, whereas in our ODE model $R$ is given by the $z$-coordinate of $E_\mathrm{max}$.}
  \label{fig:vRn_relations_phtotoi}
\end{figure}

Furthermore, Naidis uses a different definition of the radius as said in the figure caption. 
Despite all these differences, figure~\ref{fig:vRn_relations_phtotoi} shows that Naidis' approximation here fits our more systematic results quite well
\ue{for air under normal conditions}.

\newpage

\section*{References}
\bibliographystyle{iopart-num}
\bibliography{refs}

\end{document}